# Observation of Selective Isotope Effect in the Ultraviolet excitation of $N_2$: A Computational Study

B. H. Muskatel[1], F. Remacle[1,2], Mark. H. Thiemens[3], R. D. Levine[1,4]

Abstract

Isotope effects associated with gas phase $N_2$ photolysis are used to interpret Martian atmospheric evolution, icy satellite atmospheric chemistry and meteorite isotopic anomalies from nebular $N_2$ photochemistry. To interpret observations at the highest level, fundamental understanding of the precise wavelength dependency of the process must be known. In this paper VUV isotopic photodissociation effects are calculated as a function of wavelength at different wavelength slices in the 12.5-15 eV range. A very strong wavelength dependence is observed, which is significant for experiments. An observable effect is possible for the width of the beam profile at the advanced light source, ALS that may produce sufficient photolysis product for high precision isotopic analysis. A significantly more pronounced effect is predicted for a beam narrower by a factor of four providing a potential experimental test of the model. The spectrum is computed *ab initio*. It manifests two physical mechanisms for the isotope effect and they can be discriminated using a narrow beam. The fractionation is larger for the rarer heaviest isotopomer $^{15}N^{15}N$ and half as large for $^{15}N^{14}N$.

[1] The Fritz Haber Research Center, The Hebrew University, Jerusalem, Israel 91904
[2] Director of Research, FNRS. Département de Chimie, B6c, Université de Liège, B4000 Liège, Belgium
[3] Department of Chemistry and Biochemistry, University of California San Diego, La Jolla, CA 92093
[4] Department of Chemistry and Biochemistry and Crump Institute for Molecular Imaging and Department of Molecular and Medical Pharmacology, University of California, Los Angeles, CA 90095

The solar spectrum in the UV and VUV is highly structured with peaks, such as the familiar Hydrogen Lyman series (1-3). The interaction of solar or interstellar light with gas phase molecules produces photolysis and an associated chemical reaction sequence. The isotope effect associated with this photolysis and concomitant isotopic observations permits interpretation of a wide variety of processes. Martian atmospheric chemistry and evolution and interstellar molecular cloud chemistry have been studied on the basis of isotopic observations and meteorite $^{15}N$ enrichments have been suggested as arising from nebular $N_2$ photolysis (e.g. the review paper (4)) The chemistry of small satellite's such as Saturn's moon, Titan, has also been studied using nitrogen isotopic photochemical profiles (5). In each instance, thorough resolution of the isotopic fractionation processes is requisite, especially high resolution wavelength effects, to model and interpret the data.

In applications in the atmosphere and especially early solar system photochemistry, the high resolution energy dependency may become amplified. When solar light propagates through a medium, those solar peaks that are resonant with absorbing transitions will be preferentially depleted. The composition of the solar light will therefore vary along the propagation axis. This is the well known isotopic shielding effect. The role of this effect in preferential absorption of solar radiation by molecules of different isotopic composition, continues to be a subject of active research and debate, particularly for CO (5-12). An important step in enhancing understanding of the role of this process in astronomical situations is providing quantitative details of the preferential absorption of different isotopomers in an optically thin layer where all molecules experience the same light spectrum. This is attainable by reducing the path length such that opacity effects are negligible. As a contribution to understanding isotope effects associated with dissociation we report the computed frequency dependent isotopic fractionation. This high resolution wavelength examination is significant to explore giving the high degree of light structuring in natural systems. Furthermore, the calculations are carried out for conditions that allow an experimental checking. In much of the relevant UV spectral region, it is difficult to test the wave length dependence of the isotopic effect under laboratory conditions .The issue is not in the tenability of the light source, of which there are many, the limit is that one must produce sufficient high precision isotopic to test the observed model. At present, at the part per thousand sensitivity measurement ability of $^{15}N/^{14}N$ this is achievable through off line isotope ratio measurements. This protocol for example has been achieved by (10) for CO at the Advanced Light Source facility at Berkeley. As such, synchrotron radiation is an excellent option for model testing, though future laboratory based analysis may be used and the present calculations will be important for development of such techniques. In this paper, the calculations of wavelength dependent photolysis are performed using as a VUV light source comparable to the beam energies and widths available from the advanced light source, ALS, at the Lawrence Berkeley Laboratory. This beam has a width of ~0.5 eV at FWHM (see figure 1) (12) and we here use the same beam profile, and one with a smaller width at all the frequencies in the energy range of 12.5 to 15 eV. The beam width at the ALS is controlled by an undulator and may in fact be tightened to 50-100 mev (13) but with a restriction in photon fluence by a factor of ~1000. The FWHM chosen is limited by the number of photons required to produce sufficient product to collect and isotopically analyze off-line by state of the art isotope ratio mass spectrometry at the isotopic precision required to test the model (10). The present results are of significance because they not only provide further details on the high energy sensitivity of isotopic photodissociation, they also provide guidance for future synchrotron and laboratory based experiments. To test the existing model (14) several wavelengths for experimental observation must be selected and are critical for the design of the experiment.

When there is an almost negligible reduction in the intensity of the incident light the absorption rate for the molecules that absorb light is the product of the molecular absorption cross section and the incident light intensity. Both depend on the frequency and for an incoherent source integration over the relevant range in frequency is required. Unlike the solar spectrum, in the lower part of the energy range of interest the beam profile of the ALS that is planned to be used in the experiment (12) is relatively broad, as shown in figure 1. The width is set by the need for providing a sufficient number of photons to produce ample product for a off line high precision isotope ratio analysis It is roughly 0.5 eV FWHM (10), so that for a molecule such as $N_2$, typically more than one vibrational state is within the range in frequency where the beam intensity is high as can be seen in figure 1. In our calculations, the effect of energy and beam width are quantitatively assessed and reported. The short summary is that reducing the beam width significantly enhances the magnitude of the observable effect.

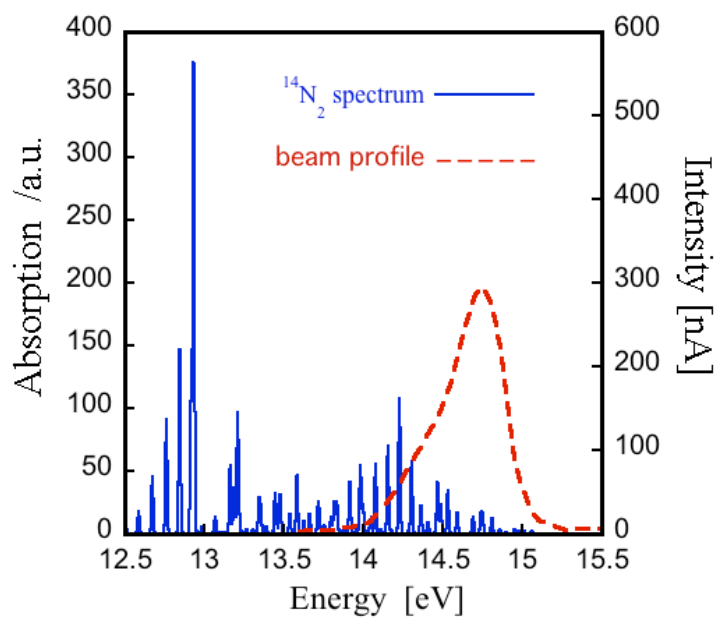

Figure 1. The *ab initio* computed spectrum of $^{14}N^{14}N$ vs. the frequency in the energy range 12.5-15.5 eV and the beam profile, dashed line, available from the ALS, (12) at an energy of 14.75 eV. The computed rate of absorption is proportional to the area under the product of both. We generate the rate of absorption $J$ vs. energy by varying the central frequency of the beam while maintaining its shape constant.

We have recently analyzed the absorption spectrum of different isotopomers of diNitrogen (14). Another very recent study is (15). Other recent high-level studies include works include (16,17). It is qualitatively and even semi quantitatively reasonable to distinguish between two sources of an isotope effect that can lead to isotopic selectivity in light absorption. One specific effect is the very long recognized mass-dependent shift of vibrational spectral lines. This small but finite shift is amply sufficient to allow for an almost complete spectral separation whereby one isotopomer absorbs and the other not. There are different options for scavenging those molecules that absorbed and in practice this matters in terms of the efficiency of isotope separation. In principle with a frequency stabilized light source one can use this isotope effect to achieve an almost complete isotope separation. The mass-dependent isotope energy shift is inherently included in our computation but it cannot be the entire origin of the experimental effect that we look for because the shift is far smaller than the width of the light beam from the ALS. There are even smaller shifts of the vibrational energy states such as due to coupling to the triplet states, see e.g., (17). These couplings are however weak and the resulting shifts are insignificant compared to the energy resolution that is discussed here.

There exists however a second isotope effect that affects the relevant absorption spectrum. Simplistically the effect modifies the spectral intensities (14) and is complementary to the familiar effect of a shift in the spectral position resulting from isotopic substitution. Even in the most elementary description of a diatomic molecule there is an isotopic shift of the intensity due to a change in the Franck Condon factors. The effect is larger in diHydrogen and smaller in heavier diatomics due to the far smaller fractional change in the mass. This effect on spectral intensities is systematic and goes together with the shift in frequency. It is not this effect that we highlight. Rather, we discuss an accidental effect due to the coupling of different excited electronic diabatic states. In the VUV range these are the excited valence and Rydberg states (18,19). Valence excitation is accompanied by a significant weakening of the N-N bond. The potential energy curve, (see (20) for accurate computations) is then likely to cross the potential of the more strongly bound Rydberg states. This crossing gives rise to localized perturbations in the spectrum and it is in the region of these perturbations that we expect the second concurrent isotope effect. Its origin is the accidental mixing of the valence excited states that carry a high oscillator strength, and the Rydberg states. It is localized in energy and occurs when a vibrational level is perturbed while the levels above and below it in the ladder remain largely unmixed (21,22). The energy region of intensity scrambling is therefore of the order of a vibrational spacing or more. This second effect can thus be experimentally significant even for a relatively broad in frequency light beam.

We compute the spectrum as follows. We first generate a basis of 250 excited vibronic states. In the energy range, 12.5-15 eV, of interest, it is a practically numerically complete basis for states that are one photon accessible from the ground state. Basis states carry a label of $\Pi$ or $\Sigma$ electronic symmetry that is an exact symmetry at our level of approximation. One valence excited and two Rydberg diabatic states are identified for each symmetry (18). The potential energy curves for each diabatic state are those computed by Spelsberg and Meyer in ref (20). These are smoothly varying potentials and we compute 45 vibrational eigenstates for each curve except the lowest lying and shallow b valence excited state for which 25 states are computed below the continuum. The label of which isotopomer enters at this stage because it is needed to specify the kinetic energy operator. A Hamiltonian matrix in the vibronic basis is diagonalized. The Hamiltonian matrix has a block form. Along the diagonal are block diagonal matrices that are the vibrational spectrum of a given diabatic electronic state. These matrices are diagonal because the different vibrational states of the same electronic state are orthogonal. The diagonal matrices are coupled by off diagonal matrices that are the vibronic couplings. These are computed as vibrational matrix elements of the electronic coupling of different diabatic states. The Hamiltonian matrix has two large and uncoupled blocks corresponding to the two electronic symmetries that can couple to the ground state by one photon excitation. The oscillator strength, $\sigma$, to the resulting eigenstates is computed from the matrix representation of the transition dipole operator from the vibrations of the ground electronic state, where 45 states are included. The spectrum as a function of frequency is given by these oscillator strengths, each with a room temperature rotational envelope. This more traditional method of computing the spectrum has been verified against the time dependent approach described in (14).

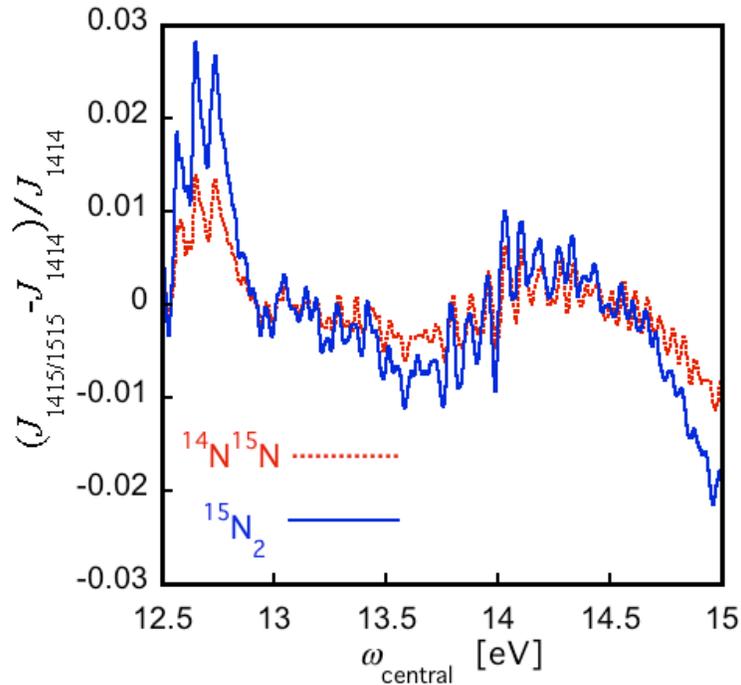

Figure 2. Predicted excess isotopic fractionation for the two rarer isotopomers of $N_2$ vs. the central beam frequency $\omega$ for a beam with a profile expected from the advanced light source, as reported in (12). $J_\omega = \int dv\, I_\omega(v)\sigma(v)$ where $I$ is the beam flux and $\sigma$ is the oscillator strength for the particular isotopomer.

Figure 2 shows the results when integrating over the computed spectrum multiplied by the expected beam profile. Shown is the rate of absorption $J$ of the two rarer isotopomers relative to that of the abundant, $^{14}N^{14}N$, isotopomer vs. the central frequency $\omega$ of the beam. The fractionation rises above 10 ‰. Despite the (0.5eV) width of the ALS beam, cf. figure 1, it is seen that the effect does vary with the central frequency of the beam, reflecting different states residing within and out of the beam center. The fractionation is not large, about 10 per mil at selected wavelengths for $^{14}N^{15}N$, but it is within experimental feasibility. (For nitrogen typical errors associated with measurement are of the order of 0.03 per mil).

A two orders of magnitude larger effect is reported in figure 3 where the beam width is reduced by a factor of ten. Reducing the width by a factor of four suffices for an order of magnitude larger isotope effect. Larger isotopic enhancements at higher wavelength resolution will occur. But testing is limited by the amount of product that is produced by photolysis and separated for off-line isotope ratio analysis. Advances in this capability (13) will provide a tighter examination of the model.

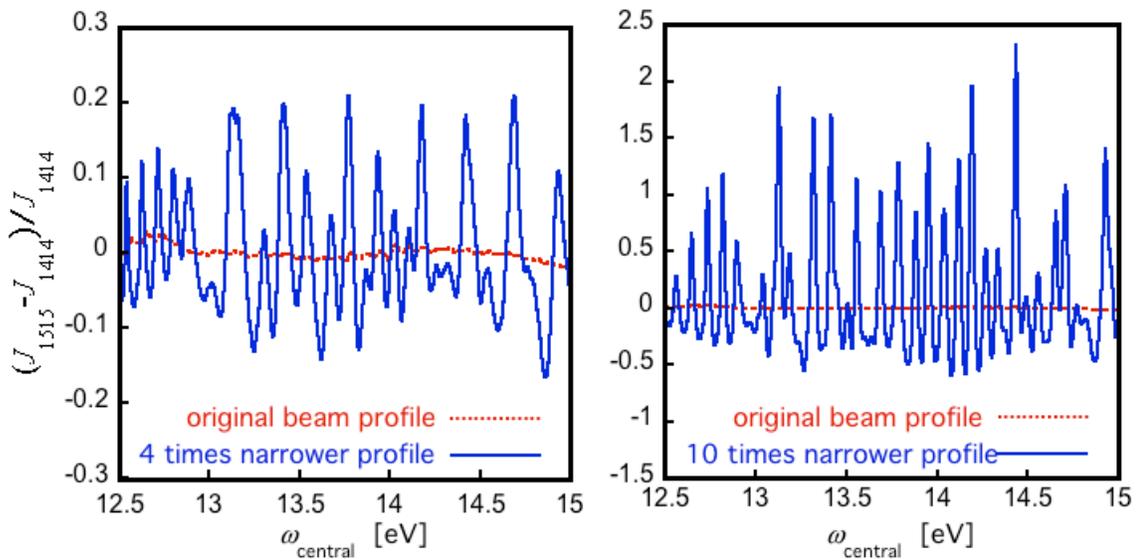

Figure 3. The very significant enhancement of the isotopic fractionation predicted for narrower beams, left panel: width of 0.13eV FWHM, right panel; width of 0.05eV FWHM. Note the change in the scale of the ordinate between the left and right panels and that the scale in figure 2 is one order of magnitude smaller. The isotope effect is shown only for the rarest isotopomer $^{15}N^{15}N$ (blue online). The predicted effect for the more common, $^{14}N^{15}N$ isotopomer, is about half as large. The effect computed for the original beam profile shown in figure 2 is here plotted as a dotted line.

Figure 3 shows the results predicted for isotopic fractionation when the beam width is ¼ as large, left panel and 1/10 as large, right panel, as for figure 1. There is a dramatic enhancement in selectivity suggesting that experiments with a narrower slit will be the most sensitive for testing the model of (14). Experimental choice of wavelengths associated with the peaks and valleys in the rate will afford the largest range. When the beam profile is reduced by a factor of 10, the isotopic structure is even more observable at a wavelength of approximately 14.4 eV there is a near full factor of two enrichment in the heavy isotopes of nitrogen compared to nearby surrounding wavelengths. This dramatically shows the tight structuring of the isotopic dependency. In applications in natural environments where there are strong source wavelength variations, optical shielding (both self and non self), and dust effects, this information is critical in interpretation of the observational data. Furthermore, the structure in figure 3 suggests that photochemical experiments at higher energy resolution will be an excellent test of the present calculations.

In conclusion, using an *ab initio* computed spectrum of $N_2$ in the far UV we show that it is possible to use the ALS or a similar tunable light source (of suitable photon flux for isotopic measurement) to demonstrate isotopic fractionation in the absorption and to test photochemical isotope effect models. The origin of the effect is due to both isotopic shift of position and of intensity of the absorption lines. The latter is due primarily to the localized and accidental mixing of valence excited and Rydberg states.

Acknowledgments: We thank T.J. Martinez, J. Troe and G. J. Wasserburg for their comments on our work. MHT acknowledges numerous helpful discussions with S. Chakraborty, G. Dominguez and R. Shaheen and Antra. Details of the ALS operation by S. Chakraborty were most helpful. The work of BHM, FR and RDL is supported by the James Franck Program for Laser-Matter Interaction. The work of RDL and FR and the computational facilities for this project is also supported by the EC FET Nano-ICT project MOLOC. Support for MHT was provided by NASA Origins of Solar Systems and Cosmochemistry programs.